\documentclass[prl,a4paper,twocolumn,showpacs]{revtex4}
\usepackage{graphicx,amsmath}
\newcommand{\be}{\begin{equation}}
\newcommand{\ee}{\end{equation}}
\newcommand{\bea}{\begin{eqnarray}}
\newcommand{\eea}{\end{eqnarray}}
\newcommand{\ba}{\begin{array}}
\newcommand{\ea}{\end{array}}
\newcommand{\ben}{\begin{enumerate}}
\newcommand{\een}{\end{enumerate}}
\newcommand{\bei}{\begin{itemize}}
\newcommand{\eei}{\end{itemize}}

\newcommand{\f}[2]{\frac{#1}{#2}}

\newcommand{\om}{\omega}

\newcommand{\bd}{b^\dagger}

\newcommand{\sgp}{\sigma_+}
\newcommand{\sgm}{\sigma_-}
\newcommand{\sgz}{\sigma_z}

\newcommand{\erfc}{\text{erfc}}

\begin{document}
\title{Atom in a coherently controlled  squeezed vacuum}
\author{Itay Rabinak, Eran Ginossar and Shimon Levit}
\affiliation{Department of Condensed Matter Physics, The Weizmann Institute of Science, Rehovot 76100, Israel}
\email{itay.rabinak@weizmann.ac.il}
\date{\today}
\begin{abstract}
A broadband squeezed vacuum photon field is characterized by a  complex squeezing function.
We show that by controlling the wavelength dependence of its phase  it is possible to change the dynamics of the atomic polarization interacting with the squeezed vacuum.
Such a phase modulation  effectively produces a finite range  temporal interaction kernel  between the two quadratures of the atomic polarization yielding the change in the decay rates as well as the appearance of additional oscillation frequencies.
We show that decay rates slower than the spontaneous decay rate can be achieved even for a squeezed bath in the classic regime.
For linear and quadratic phase modulations the power spectrum of the scattered light exhibits narrowing of the central peak due to the  modified decay rates.
For strong phase modulations  side lobes appear symmetrically around the central peak reflecting additional oscillation frequencies.
\end{abstract}
\pacs{78.67.De, 42.50.Dv, 42.55.Sa, 42.50.Lc}

\maketitle

The effect of the interaction of an atom with a squeezed light had
been studied extensively. Gardiner \cite{Gardiner86} has
considered the behavior of a two-level atom  damped by an
infinite bandwidth squeezed vacuum. He showed that the two quadratures of the atomic polarization decay at
different rates. Decay rates smaller than the decay rates of
spontaneous emission can be achieved for non-classical squeezing. The presence of
the two decay rates modify the fluorescence spectrum, Ref.\ \cite{Carmichael87}.
In Refs.\ \cite{In_Out_put1,In_Out_put2,finit_band_compl} the interaction of a finite bandwidth
squeezed vacuum with a two level atom was investigated.
Review of these and related studies is found  in Ref.\ \cite{atom_sqz_review}.
Recently these works were extended in Ref.\ \cite{ginossar}  to include
interactions with  semiconductor microstructures.

In the field of quantum coherent control 
pulse shapers \cite{pulse_shape} were used to attain prescribed phase modulation of a down converted light in
order  to control two photon absorption. It was shown
\cite{silberg} that pulses can be shaped in a way that
will stretch them temporally affecting the transition
probability \cite{Nirit}.
Experiments \cite{nonclass,nonclass2} have also been performed on
two photon absorption with coherent, narrow band down-converted
light, demonstrating nonclassical features which appear at very
low powers \cite{non_exp,non_exp2,non_exp3} and result from
time and energy correlations (entanglement) between the
down-converted photon pairs.

In this work we wish to investigate the effect of controlling and
modulating  the relative phase of the modes of a squeezed
reservoir interacting with an atom.  This can be easily done  by a pulse shaper arrangement
as shown in  Refs.\ \cite{silberg,dayan}.  We will demonstrate
that  by controlling in this manner the phases of the correlations in the squeezed reservoir  it is possible to control
the dynamics of atomic polarization and in particular to further reduce its decay rates.

We will use the standard model to describe the interaction of a
two-level atom with a broadband radiation field.  The atom is assumed to be coupled to
a one dimensional set of radiation modes.  The Hamiltonian
is given in the electric-dipole and rotating-wave approximations
by $H=H_0+H_I$ as (we take $\hbar=1$): \bea
    &&H_0=\omega_a \sigma_z + \sum_q \omega_q b^{\dag}_q b_q \nonumber \\
    &&H_I=\Gamma \sigma_+ + \sigma_-\Gamma^{\dag} \,\,  , \,\, \Gamma = \sum_q g_q b_q.
\eea
The pseudospins $ \sigma_+, \sigma_-$and $\sigma_z$ describe the atom, $\omega_a$ is the atomic resonance frequency,
$b_q$ and $b^{\dag}_q$  describe radiation  modes with  wave vector $q$  and frequency $\omega_q$ and $g_q$ are the mode-atom couplings.

We assume that the radiation acts as a reservoir with correlations of a two mode broadband
squeezed vacuum~\cite{Gardiner86,In_Out_put1}
\bea \label{field-corrs}
\label{def_m}
&&\langle \bd_q b_{ q'}\rangle={N}(\om_q)\delta_{q,q'} \nonumber \\
&&\langle \bd_q b^{\dag}_{
q'}\rangle={M}(\om_q)\delta_{2q_{0}-q, q'} \\
\nonumber &&\langle b_q \rangle=\langle b^{\dag}_q \rangle=0,\\
\nonumber
\eea
where ${N}(\om_q)$ describes the average occupations
of photonic modes while the magnitude $|{M}(\om_q)|$ gives the
squeezing strength of mode pairs centered around the frequency
$\om_0\equiv\om(q_0)$. The phase of ${M}(\om_q)$ describes the ``direction'' of
squeezing in the phase space of squeezed mode pairs. In
the common method of generating squeezed vacuum radiation  by non-linear down conversion the phase of
 ${M}(\om_q)$ is constant over the frequency range. Letting the radiation pass through a pulse shaper makes it possible to change this phase into
  a prescribed function of $\omega_q$,  cf., Refs.\ \cite{silberg,dayan}.

 We shall for simplicity  assume that $ {N},\, |{M}|, g_q$ are constants within
a bandwidth $\om_0\pm B/2$ and that  $B$ is $\ll \omega_0$ but is much larger than any other
frequency in the system.

Using  the equations of motion
\bea
\label{eom_basic}
\dot{\sigma}_+ &=& -i[\sgp,H] = i\om_a \sgp - 2i\Gamma^{\dag}  \sgz  \nonumber  \\
\dot{\sigma}_z &=& -i[\sgz,H] = -i\Gamma \sgp + i\Gamma^{\dag} \sgm  \\
\dot{b}_q &=& -i[b_q,H] = -i\om_q b_q -ig^*_q \sgm \nonumber
\eea
and transforming them to the rotating frame of the laser frequency,
we integrate over time the equations for $\sigma_z$ and $b_q$  and
substitute the result back into the equation for $\sigma_+$.
We then average  the resulting equation over the initial state and  assume that the bath and the atom are (approximately) decorrelated, i.e.
that the photon atom correlators factorize at all times e.g.  $ \langle b_q(t) \sgp(t')\rangle \approx \langle b_q(t)\rangle \langle \sgp(t') \rangle $,  etc.
The decorrelation assumption is valid in the regime of  weak coupling between
the system and the photon bath. Using it we obtain a closed equation for the atomic polarizations  $\langle \sigma_{\pm} \rangle$
\bea
\label{avg_pol_der}
   \frac{d}{dt} \langle \sigma_+ \rangle &=& \left(i(\om_a - \om_0) - \f{\gamma}{2}\right)\langle \sgp \rangle  \\
    &-& 2  \int^t_0 dt' \{ \sum_{q}
     |g_q|^2 {N}(\om_q)e^{i (\om_q-\om_0)(t-t')} \langle\sgp(t')\rangle \nonumber \\
    &-& g^*_q g^*_{Q}{M}(\om_q) e^{i (\om_q -\om_0)(t-t')} \langle\sgm(t')\rangle \} \nonumber ,
\eea
where $ \om_Q = 2\om_0 - \om_q $,  $\gamma = \rho(\om) |g_q|^2$
is the vacuum atomic decay rate and $\rho(\om)$ is the density of the radiation modes. We assume that $\rho(\om)$
is flat over the bandwidth B.

We now transform the sums over $q$ into integrals. We define the following parameters
\be
\begin{split}
    \gamma\mathcal{N} &=  \rho(\om) |g_q|^2 N(\om_q) \\
    \gamma\mathcal{M} &= \rho (\om)  g^*_q g^*_{Q} |{M}(\om_q)| ,
\end{split}
\ee
and the following function
\be
    k(t-t') = \f{1}{\pi}\int_{-B/2}^{B/2}d\om e^{if(\om)} e^{i \om(t-t')}.
\ee where $f(\om)$ is the phase of $M(\om) = |M|e^{if(\om-\om_0)}$. Note that $f(\om)=f(-\om)$. Without loss of generality we can take
$f(0)=0$ using the freedom to absorb its non-zero value in the phase of $\mathcal{M}$.

With this notation  the equation for the atomic polarizations becomes
\be
\label{fin_pol_dif}
  \frac{d}{dt} \langle \sigma_+ \rangle= \left(i\delta - \gamma(\mathcal{N}+ \f{1}{2})\right)\langle \sgp \rangle
    + \gamma\mathcal{M} \int^t_0 dt'  k(t-t')  \langle \sgm(t')\rangle,
\ee
where we  defined the detuning $\delta = \om_a - \om_0$ .

The phase modulation of the squeezing parameters leads to a finite range memory kernel in Eq. (\ref{fin_pol_dif}) which couples atomic polarizations
to their complex conjugates at earlier times. This non-Markovian dynamics  is still linear due to the atom-bath decoupling assumption. One can envisage
 situations in which the non-Markoian dynamics  may lead to the breakdown of this assumption but we will not consider
such cases here.

To analyze the controlled memory effects we apply the Laplace transform to Eq. (\ref{fin_pol_dif})
and  obtain
\begin{widetext}
\bea
\label{sol_pol_s}
\left(
      \begin{array}{c}
        \langle \tilde{\sigma}_- (s)\rangle\\
        \langle \tilde{\sigma}_+(s) \rangle\\
      \end{array}
    \right)= \f{\left(
      \begin{array}{cc}
        s+i\delta + \gamma(\mathcal{N}+ 1/2) & \gamma\mathcal{M}\tilde{k}(s) \\
        \gamma\mathcal{M}^*\tilde{k}^*(s^*) & s-i\delta + \gamma(\mathcal{N}+ 1/2) \\
      \end{array}
    \right)
    \left(
      \begin{array}{c}
       \langle \sgm(0) \rangle \\
       \langle \sgp(0)\rangle \\
      \end{array}
    \right)}
    { (s+ \gamma(\mathcal{N}+ 1/2))^2 +\delta^2 - |\mathcal{M}|^2\tilde{k}(s)\tilde{k}^*(s^*)  }
\eea
\end{widetext}
where we assumed  non-zero initial conditions at $t=0$ for $\langle \sigma_{\pm}(t) \rangle$ and
denoted $\langle \tilde{\sigma}_{\pm}(s) \rangle,  \tilde{k}(s)$  the Laplace transform of the polarizations and the kernel function respectively.
The non-zero $\langle \sigma_{\pm}(0) \rangle$ is the easiest situation to analyze although it can only be realized with a specially
designed initial pulse applied before letting the squeezed reservoir interact with the atom.  We will later discuss the implications for the fluorescence spectrum.
In the following we will assume for simplicity the exact resonance case $\delta=0$.

\begin{figure}
\includegraphics[scale=0.58]{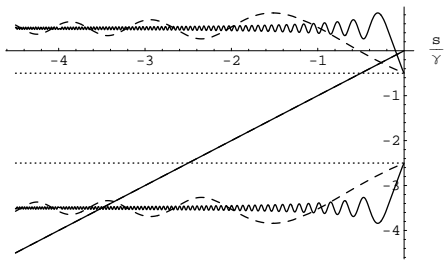}
\caption{Graphical solution of Eq. (\ref{plot_sol_s}) with $\mathcal{N}=\mathcal{M} =1 $ for the quadratic phase modulation, Eq. (\ref{quad ker}).
 The wavy lines show the right hand side of the equation for $T\gamma=5$ (solid line) and $T\gamma=1$ (dashed line) while the horizontal dotted lines are for $T=0$
 -  the standard Markov case. \label{roots}}
\end{figure}

The poles of the Laplace transform  are the solutions of
\be \label{plot_sol_s}
    s/\gamma= - \mathcal{N} - 1/2  \pm |\mathcal{M}|\sqrt{\tilde{k}(s)\,\tilde{k}^*(s^*)}.
\ee
For the unmodulated phase $f(\om)=0$  we obtain  the usual Markov result  $k(t)=\delta (t)$,  $\tilde{k}(s) = 1,$ in which case
$s_{\pm} = \gamma(\mathcal{N}+ 1/2  \pm \gamma|\mathcal{M}|)$, representing
the splitting of the decay rate into the fast and the slow components under the influence of the squeezing, Ref.\ \cite{Gardiner86}.
For the modulated phase it is convenient to consider the graphic solution, cf. Fig.\ref{roots}. The symmetry of $f(\omega)$ implies that $\tilde{k}(0)=1$.
 Therefore for analytic $\tilde{k}(s)$ to the first order
 in $s$, $\tilde{k}(s) \approx  1 -s\int_0^{\infty}tk(t)dt\equiv1-sk_1$. The right hand side of  Eq.(\ref{plot_sol_s}) reduces to
    $- \mathcal{N} - 1/2 \pm |\mathcal{M}|(1-s {\rm Re} k_1)$
and if ${\rm Re} k_1 >0$ then  the slow component of the decay rate
\be
\label{s min dec}
    s_{-}= -\gamma (1 - \gamma| \mathcal{M}|{\rm Re}k_1)(\mathcal{N} + 1/2 - | \mathcal{M}|)+\dots
\ee
can become even slower than in the Markov case.

We wish to remark that  in addition to the real poles, the complex poles  of Eq. (\ref{sol_pol_s})  in the ${\rm Re}s<0$ half plane also  play a role
in determining the dynamics of the atomic polarizations. Their imaginary parts represent additional oscillation frequencies. If their real part  is smaller than the real poles they may
dominate the  long time decay.  Finally possible cuts may also make a contribution.  We will discuss  such contributions
in the examples below.

We consider two  concrete examples of the  phase modulation.
The simplest case to calculate is the  \emph{quadratic phase}
\be
\label{quadratic}
f(\om) = T^2 \om^2,
\ee
where $T$ is a real positive parameter. The resulting Laplace transformed memory  kernel in the $B\to\infty$ limit is
\be
\label{quad ker}
\tilde{k}(s) =  e^{-i T^2 s^2} \erfc(\sqrt{-i}T s).
\ee
The solution of  (\ref{plot_sol_s}) with this $\tilde{k}(s)$ is shown graphically  by plotting
the two sides of this equation  in Fig.\ref{roots}.  One sees the increase of the fast and the decrease of the slow decay rates caused by the phase modulation.
In this example $\gamma{\rm Re}k_1$  in (\ref{s min dec})  is ${\sqrt{\frac{2}{\pi }}} T\gamma$.

\begin{figure}
\includegraphics[scale=0.58]{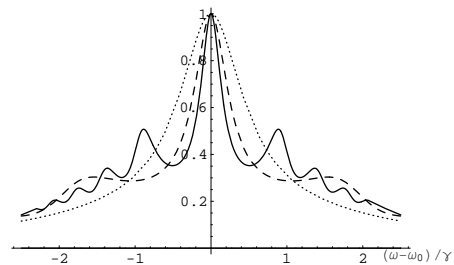}
\caption{The fluorescence spectrum for the quadratic phase modulation, Eq. (\ref{quadratic}) and  $\mathcal{N}=\mathcal{M} =1$.
Solid, dashed and dotted  lines are for $T\gamma=5$, $T\gamma=1$ and $T=0$ (the Markov case) respectively. \label{sqrd_MeqN}}
\end{figure}

For $\tilde{k}(s)$ given by Eq. (\ref{quad ker}) the structure of Eq. (\ref{sol_pol_s}) is much richer than that implied by the real poles. One finds in addition numerous complex poles, the
positions of which depend on the parameters $\mathcal{N}$, $\mathcal{M}$, $T$ and $\gamma$.
 Their influence can be seen in Fig. \ref{sqrd_MeqN} where we plot $\langle \tilde{\sigma}_-  (s = i\om) \rangle/\langle \tilde{\sigma}_-  (s = i\om_0) \rangle$.
 Together with the significant narrowing of the central part of the peak relative
to the Markov case $T=0$ one observes the appearance of the side lobes which reflect a complicated pole structure in the  $Res<0$ half plane.

 The quantity plotted
in Fig. \ref{sqrd_MeqN} can be related to the fluorescence spectrum of light emitted by the atom into empty modes of the radiation field.
This spectrum is given by
\be
    S(\om) = \f{\gamma}{2\pi} {\rm Re}\bigg\{  \int_{0}^{\infty} \text{d}\tau \, e^{-i\,\om \tau}
    \langle \sgp(t_0+\tau)  \sgm(t_0) \rangle_{ss}\bigg\}.
\ee
where the subscript $ss$ denotes steady state. This expression  involves the average  of the two time product of polarization operators $\langle \sgp(t_0+\tau)  \sgm(t_0) \rangle$.
 It is not difficult to show that for $t_0=0$  and $t_0+\tau=t$ this average satisfies the same
dynamical equation (\ref{fin_pol_dif}) as a single time average $\langle \sgp(t)\rangle$ provided that one  uses the atom - bath  decorrelation assumption. Irrespectively of the initial condition
the system relaxes to a steady state regime.  One can therefore replace the initial time, i.e. the low limit $t=0$ in the integral  in  Eq. (\ref{fin_pol_dif}) by $t_0$ provided it is
 chosen after the steady state is reached. The initial condition is then  given by $\langle \sgp(t_0)  \sgm(t_0) \rangle_{ss}=1/2 + \langle\sigma_z(t_0)\rangle_{ss}$.
  The Laplace transformed solution  of this set  is given by (\ref{sol_pol_s})
with $\langle  \widetilde{\sigma_{\pm} \sgm} \rangle (s)$ replacing $\tilde{\sigma}_{\pm}(s)$ in the left hand side and $\langle \sigma_{\pm}(0) \rangle$ replaced by
\be
\begin{split}
     \langle \sgm(0) \rangle &\rightarrow \langle \sgp(t_0)  \sgm(t_0) \rangle_{ss} =\f{1}{2} - \f{1/2}{2N + 1}  , \\
     \langle \sgp(0)\rangle &\rightarrow \langle \sgp(t_0)  \sgp(t_0) \rangle_{ss}=0.
\end{split}
\ee
The fluorescence spectrum is given by $S(\om) = \f{\gamma}{2\pi} {\rm Re}\big\{ \langle  \widetilde{\sgp \sgm} \rangle (i \om)  \big\}$,  the normalized form of which is
the quantity plotted in Fig. \ref{sqrd_MeqN}.

 We note that the discussion above essentially means that under the set of the adopted assumptions the quantum regression theorem can be applied
  despite the finite memory effects induced by the phase modulation.

\begin{figure}
\includegraphics[scale=0.58]{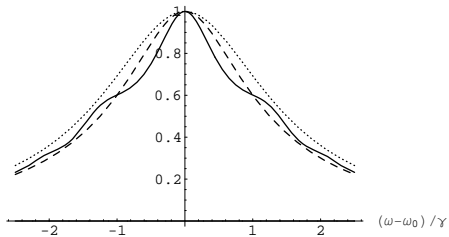}
\caption{ Sensitivity of the fluorescence spectrum to the degree of squeezing.
$T\gamma=2, \mathcal{M}=0.5,$ (solid line), $T=0, \mathcal{M}=0.5,$ (dashed)
and $\mathcal{M} =0$ (dotted), white noise reservoir. For all graphs $\mathcal{N}=1$.
\label{sqrd_2MeqN}}
\end{figure}
In Fig. \ref{sqrd_2MeqN} we show how the phase modulation increases the sensitivity to squeezing. The graphs represent three possible reservoirs:
squeezed phase modulated, squeezed unmodulated and white noise reservoir. One observes that the narrowing of the central peak
is present even  for the reduced values of $M$ provided a strong phase modulation is applied.

\begin{figure}
\includegraphics[scale=0.58]{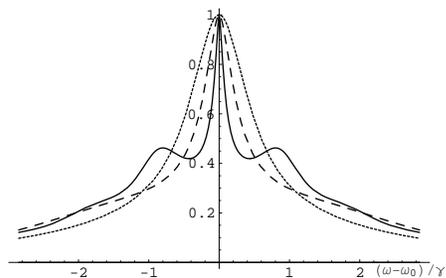}
\caption{Linear phase modulation (\ref{lin mod}) for $\mathcal{N}=1,\mathcal{M} =1 $ and
$T\gamma=5$ (solid line),  $T\gamma=1$  (dashed)
$T=0,$ (doted),  the normal Markov case. \label{abs_MeqN}}
\end{figure}

We will now briefly discuss  another example, the \emph{linear phase modulation}
\be
\label{lin mod}
f(\om) = T |\om|,
\ee
where $T$ is a real positive parameter.
In this case $\tilde{k}(s)$ is multivalued with $s=0$ as a branch point. The  pole structure is now more involved and should be
discussed together with the branch cuts in the complex $Res<0$ half plane.  In particular the discussion following Eq. (\ref{plot_sol_s}) should
be modified. We will not discuss it  here  but rather present in Fig. \ref{abs_MeqN} the quantity $\langle \tilde{\sigma}_-  (s = i\om) \rangle/\langle \tilde{\sigma}_-  (s = i\om_0) \rangle$
 for this phase modulation. As discussed above this quantity represents the  fluorescence spectrum.  We observe features similar to those found in the quadratic modulation case and in particular
 the narrowing of the spectrum around the central frequency and the development of the side lobes.  From an experimental point of view the advantages of the
 linear modulation is the relative ease of achieving it in practice.

It is a pleasure to acknowledge valuable discussions with Y. B. Levinson.


\begin{thebibliography}{10}

\bibitem{Gardiner86}
C.~W. Gardiner, Phys. Rev. Lett. {\bf 56},  1917  (1986).

\bibitem{Carmichael87}
H.~J. Carmichael, A.~S. Lane, and D.~F. Walls, Phys. Rev. Lett. {\bf 58},  2539
   (1987).

\bibitem{In_Out_put1}
C.~W. Gardiner and M.~J. Collett, Phys. Rev. A {\bf 31},  3761  (1985).

\bibitem{In_Out_put2}
C.~W. {Gardiner}, A.~S. {Parkins}, and M.~J. {Collett}, J. Opt. Soc. Am. B {\bf 4}, 1683 (1987).

\bibitem{finit_band_compl}
A.~S. Parkins and C.~W. Gardiner, Phys. Rev. A {\bf 40},  3796  (1989).

\bibitem{atom_sqz_review}
D.~J. Dalton, Z. Ficek and S. Swain, J. Mod. Opt. {\bf 46},  379  (1999).

\bibitem{ginossar}
E. Ginossar and S. Levit, Phy. Rev. B {\bf 72},  075333  (2005).

\bibitem{pulse_shape}
M.~M. Wefers and K.~A. Nelson, Opt. Lett. {\bf 20},    (1995).

\bibitem{silberg}
D. Meshulach and Y. Silberberg, Nature {\bf 396},    (1998);
Phys. Rev. A {\bf 60},  1287  (1999).

\bibitem{Nirit}
N. Dudovich, B. Dayan, S.~M. Gallagher~Faeder, and Y. Silberberg, Phys. Rev.
  Lett. {\bf 86},  47  (2001).

\bibitem{nonclass}
N.~P. Georgiades {\it et~al.}, Phys. Rev. Lett. {\bf 75},  3426  (1995).

\bibitem{nonclass2}
N.~P. Georgiades, E.~S. Polzik, and H.~J. Kimble, Phys. Rev. A {\bf 55},  R1605
   (1997).

\bibitem{non_exp}
J. Gea-Banacloche, Phys. Rev. Lett. {\bf 62},  1603  (1989).

\bibitem{non_exp2}
J. Javanainen and P.~L. Gould, Phys. Rev. A {\bf 41},  5088  (1990).

\bibitem{non_exp3}
H.-B. Fei {\it et~al.}, Phys. Rev. Lett. {\bf 78},  1679  (1997).

\bibitem{dayan}
B. Dayan, A. Pe'er, A.~A. Friesem, and Y. Silberberg, Phy. Rev. Lett.
  {\bf 93},  023005  (2004).

\end{thebibliography}
\end{document}